# Electrically tunable strong coupling in a hybrid-2D excitonic metasurface for optical modulation

Tom Hoekstra[1] and Jorik van de Groep[1*]

[1] *Van der Waals-Zeeman Institute, Institute of Physics, University of Amsterdam, Amsterdam, the Netherlands*

[*] j.vandegroep@uva.nl

**Abstract**
Atomically thin semiconductors exhibit tunable exciton resonances that can be harnessed for dynamic manipulation of visible light in ultra-compact metadevices. However, the rapid nonradiative decay and dephasing of excitons at room temperature limits current active excitonic metasurfaces to few-percent efficiencies. Here, we leverage the combined merits of pristine 2D heterostructures and non-local dielectric metasurfaces to enhance the excitonic light-matter interaction, achieving strong and electrically tunable exciton-photon coupling at ambient conditions in a hybrid-2D excitonic metasurface. Using this, we realize a free-space optical modulator and experimentally demonstrate 9.9 dB of reflectance modulation. The electro-optic response, characterized by a continuous transition from strong to weak coupling, is mediated by gating-induced variations in the free carrier concentration altering the exciton's nonradiative decay rate. These results highlight how hybrid-2D excitonic metasurfaces offer novel opportunities to realize nanophotonic devices for active wavefront manipulation and optical communication.

**Introduction**
Van der Waals materials and their heterostructures have emerged as a transformative materials platform in photonics, as the layered 2D structure and lack of dangling bonds offer atomic-level control over thickness in arbitrary stacking configurations[1,2]. Combined with properties such as high refractive indices, large anisotropies, and broad tunability, the prospects of two-dimensional (2D) materials in nanophotonics research are tantalizing. In particular, monolayer 2D semiconductors stand out for their uniquely strong exciton resonances which dominate their room-temperature optical properties due to the exceptionally large binding energies arising from quantum- and dielectric confinement in the atomically thin crystal[3]. Additionally, these excitons are highly sensitive to external stimuli[4–8], with free-carrier injection emerging as an attractive tuning mechanism with potential optoelectronic applications including light detection and ranging[9], free-space optical communications[10], and analog image processing[11].

The efficient electrical tunability observed in monolayer 2D semiconductors would be an enormous asset in the field of metasurfaces. Already, these ultra-compact optical coatings provide remarkable new ways to manipulate light by harnessing the strong light-matter interactions offered by resonant metallic or dielectric nanostructures[12–16]. Yet, so far, dynamic metasurfaces have been few and far between due to the weak tunability of plasmonic and geometric resonances[10]. While a variety of active devices have already been demonstrated[17–21], they are typically hindered by small modulation depths, excessive complexity, or operation limited to the infrared spectral range. As such, it seems natural to capitalize on the large tunability of excitons in 2D materials. Indeed, tunable excitonic metasurfaces have already been created by directly carving zone plate lenses and metagratings out of monolayers[22,23]. However, their efficiencies are restricted to a few percent because of the atomic thickness as well as severe damping of the exciton transition at ambient conditions by nonradiative decay channels[4] and dephasing[24]. It is therefore essential to find ways to enhance the excitonic light-matter interaction strength.

To this end, 2D materials have been placed in the vicinity of optically resonant nanostructures to leverage their strong Purcell enhancements[25–29]. In particular, a monolayer semiconductor was recently integrated in a plasmonic metasurface to achieve 10% reflectance modulation as well as dynamic beam steering[30]. Nevertheless, the observed performance was still constrained by defects and disorder in the unprotected monolayer arising from incompatibilities with standard nanofabrication protocols[31]. At the same time, it is well-established that 2D materials can be protected in heterostructures by encapsulating them with hexagonal boron nitride (hBN). This additionally provides screening from local charge fluctuations and defects in the dielectric environment, resulting in pristine excitonic properties[32,33]. Even with relatively simple planar heterostructures, this has led to remarkable realizations of near-perfect excitonic absorption[24], near-unity reflection modulation[34,35], and active beam steering devices[36,37], although only at cryogenic temperatures. Hence, it is worthwhile to explore hybridizing heterostructures with metasurfaces[38,39] as a novel route towards active light manipulation beyond the cryogenic regime.

Here, we combine the pristine excitonic properties of a Van der Waals heterostructure with the nanophotonic field enhancement offered by a non-local metasurface to demonstrate a dynamically tunable excitonic metasurface with high efficiencies. This hybrid-2D configuration enables strong exciton-photon interactions, which we harness for free-space optical modulation by electrically manipulating excitons in monolayer tungsten disulfide ($WS_2$). By switching between the weak and strong coupling regimes, we demonstrate 9.9 dB modulation of the reflected beam at room temperature, which is a fivefold improvement over the previous record reflection contrast achieved with excitonic 2D materials[30]. Furthermore, we show that the gating-induced free carrier density governs the exciton's nonradiative decay rate, providing a continuous transition from strong to weak exciton-photon coupling. Altogether, these results demonstrate that, by leveraging the combined merits of dielectric metasurfaces and Van der Waals heterostructures, electrically tunable hybrid-2D metasurfaces hold the key to realize ultracompact optoelectronic devices for dynamic wavefront manipulation at ambient conditions.

**Results**

**Hybrid-2D optical modulator**

Figure 1a illustrates a free-space optical modulator whose performance is derived from the combined merits of monolayer 2D semiconductors and dielectric metasurfaces. The hybrid-2D modulator achieves strong modulation of the input field through the electrical tunability of exciton resonances in atomically thin tungsten disulfide ($WS_2$) combined with the electromagnetic field enhancement offered by guided-mode resonances[30,40]. Applying a voltage between the gold (Au) back-gate and the monolayer results in strong electron-doping of the $WS_2$, thereby efficiently quenching the so-called A-exciton transition[6]. To demonstrate this, we measure the excitonic photoluminescence emission in a flat reference device at room temperature and show a 108-fold reduction at $\lambda_0$=618 nm, while the charged exciton[41] peak at 635 nm remains mostly unaltered (Fig. 1b). Although electrical tuning of exciton emission is by now well-established, such efficient quenching has thus far mostly been limited to cryogenic temperatures[24,34,35].

The designed optical modulator combines several key elements to leverage the tunability of 2D excitons (Fig. 1c). First, the $WS_2$ monolayer is electrically connected to the ground electrode, but otherwise isolated through encapsulation with hexagonal boron nitride (hBN). This provides screening from local charge fluctuations and defects in the dielectric environment, resulting in a narrow exciton linewidth[33]. At the same time, the resulting Van der Waals heterostructure combined with the Au back-reflector functions as an asymmetric optical cavity to confine light inside the layer stack[24]. Second, to maximize the light-matter interaction we introduce a dielectric subwavelength grating on top of the heterostructure to enable resonant excitation of its fundamental $TE_0$ guided mode. The resulting guided mode resonance (GMR) provides a 74-fold electric field intensity enhancement and greatly increases the interaction length of light with the $WS_2$ monolayer (Fig. 1d). Finally, the back-reflector additionally serves as a local back-gate to $WS_2$ layer, where the bottom hBN layer functions as the gate dielectric. As such, the back-gate enables electrical manipulation of the carrier concentration in the monolayer. In the strongly electron-doped (n-doped) regime, the exciton is effectively quenched, and the reflectance spectrum shows a single critically coupled resonance, splitting into two strongly coupled resonances when the monolayer is brought to intrinsic doping instead (Fig. 1e). Consequently, in simulation, the device exhibits an absolute reflectance contrast of $\Delta R = 47.5\%$ ($\lambda_0 = 627$ nm) when it is switched between these two states. We quantify the modulation depth as $10\log_{10}(R_i/R_n)$, where $R_i$ and $R_n$ are the reflectance in the intrinsic and n-doped regimes, respectively (Fig. 1f). Using this definition, our modulator achieves 25 dB peak signal modulation at the operating wavelength.

**Tailoring the design for strong light-matter interactions**

Next, we outline our approach for maximizing the modulation efficiency by simultaneously designing for perfect absorption in the exciton-quenched state and strong exciton-photon coupling in the unquenched state. To this end, we use rigorous coupled-wave analysis (RCWA) simulations to optimize the dimensions of the heterostructure cavity as well as the grating's unit cell. Here, we choose to design the modulator for operation around the A-exciton wavelength, $\lambda_A = 620$ nm, with normal-incident TE-polarized light, while retaining a mirror-like response for TM polarization.

To start, we tailor the bottom hBN thickness such that the monolayer is positioned at a $\lambda/2$ separation from the back-reflector (Fig. 2a). In concept, this is reminiscent of the famous Salisbury screen, in which a thin absorber is separated $\lambda/4$ from a mirror to enhance absorption through constructive interference. In contrast, placing the WS$_2$ at $\lambda/2$ instead minimizes absorption losses in the cavity due to destructive interference at the monolayer position. Since TM-polarized light cannot couple to the TE$_0$ guided mode supported by the structure, it is predominantly reflected from the cavity in this spectral and angular range (Fig. 2b). On the other hand, TE-polarized light can pick up first-order grating momentum (equal to $2\pi/\Lambda$, where $\Lambda$ is the grating period) and thereby couple to the GMR, which enhances the interaction of light with the monolayer by confining the field inside the structure. We optimize the grating periodicity such that the GMR coupling condition is satisfied at normal incidence for a photon energy matching the exciton energy, $E_A = 2.00$ eV, while the duty cycle is tailored to achieve perfect absorption through critical coupling when the monolayer is strongly n-doped (Fig. 2c). In this regime, the exciton transition is quenched as the excess free electrons partially screen the Coulomb interaction and increase the exciton-electron scattering rate[6]. Hence, absorptive losses in the cavity reduce to those suffered evanescently in the gold and by means of sub-bandgap absorption in monolayer WS$_2$ owing to impurities, defects and edge states[42].

In our simulations, we model the doping-dependence of the A-exciton resonance by modifying its nonradiative decay rate $\gamma_{A,nr}$ (details in Methods section). Although this approximation does not account for subtler effects such as interconversion to charged excitons[41] (trions), it nevertheless captures the doping-dependent optical properties of the monolayer. In the intrinsic regime, we set the nonradiative decay to occur ten times faster than the radiative decay ($\gamma_{A,r}$), with characteristic rates $\hbar\gamma_{A,r} = 2.7$ meV and $\hbar\gamma_{A,nr} = 27$ meV chosen to match those observed in preliminary experiments. To simulate strong n-type doping, we artificially suppress the exciton by increasing its nonradiative decay rate to $\hbar\gamma_{A,nr} = 270$ meV. As shown in Fig. 2c, the resulting dispersion relation simply resembles the bare cavity dispersion with a slightly broader linewidth (Supplementary Fig. S1). Crucially, the n-doped dispersion is critically coupled such that the minimum reflectance $R_{min} \to 0\%$, and the $\Gamma$-point is located near $E_A$. As a consequence, a gap opens in the dispersion around $E_A$ when the monolayer is neutralized[43,44], indicative of strong exciton-photon interactions (Fig. 2d).

To evaluate the coupling strength, we additionally simulate the normal-incidence reflectance as function of grating period (Fig. 2e). This reveals the characteristic anti-crossing behavior associated with strong light-matter coupling, where two hybridized polariton modes emerge due to the rapid energy exchange between photons and a material resonance. The energy separation is proportional to the coupling strength $g = \sqrt{N}\mu E$, where $N$ is the number of excitons with transition dipole moment $\mu$, and $E$ is the local electric field strength[45]. At resonance, a pair of coupled oscillators is said to be strongly coupled if the mode splitting, characterized by the Rabi energy $\Omega_R$, is larger than the average decay rate of the uncoupled resonances, i.e., $\Omega_R > (\hbar\gamma_A + \hbar\gamma_{GMR})/2$. From simulations of the cavity in absence of the exciton resonance, we determine $\hbar\gamma_{GMR} = 23.3$ meV for the bare GMR, in good agreement with the experimentally estimated value of 22.9 meV. We subsequently fit the dispersion at intrinsic doping using a coupled mode theory (CMT) analysis[46,47] (see Methods for details). From this, we extract a Rabi splitting of $\Omega_R = 38.2$ meV and a corresponding coupling strength $\hbar g = 19.1$ meV, putting the system well into the regime of strong coupling (>25.2 meV). Instead, in the strongly n-doped regime (inset of Fig. 2e), the exciton decay rate is so large that $\Omega_R$ becomes purely imaginary and the gap closes, as expected for a weakly coupled system. Our metasurface leverages this tunability between strong and weak coupling to achieve >12 dB reflectance modulation in a broad wavelength range around $E_A$ (Fig. 2f). This highlights how integration of a 2D semiconductor in a hybrid metasurface configuration considerably enhances the excitonic modulation depth.

**Reflectance modulation by gate-tuning of excitons**

To demonstrate our approach experimentally, we employ a dry-transfer technique to assemble mechanically exfoliated $WS_2$ and hBN flakes on prepatterned Au electrodes[48]. We functionalize the heterostructure cavity by nanopatterning a subwavelength grating in a spin-coated layer of CSAR (chemical semi-amplified resist) using electron-beam lithography (Fig. 3a). Atomic force microscopy confirms the grating's periodicity and reveals excellent uniformity of the grating lines (Fig. 3b), while photoluminescence mapping spectroscopy (Supplementary Fig. S2) verifies that the exciton resonance is not detrimentally red-shifted or broadened during fabrication by virtue of the heterostructure encapsulation. We proceed by measuring the angle-resolved reflectance of the device via back-focal plane imaging spectroscopy (Supplementary Fig. S3). First, we apply a +25 V gate bias to induce strong n-doping of the monolayer and thereby quench the exciton resonance. As shown in Fig. 3c, the dispersion relation in this state resembles that of the bare cavity (Supplementary Fig. S4). Critically, the $\Gamma$-point is located very close to $E_A$, such that when we neutralize the monolayer at –25 V, an energy gap of $\Omega_R = 36.0$ meV opens in the dispersion due to strong exciton-photon coupling (Fig. 3d).

Before proceeding, we make three noteworthy observations about these results. First, to neutralize the $WS_2$, we reverse the bias from +25 V (electron injection) to –25 V (hole injection), indicating that the monolayer is significantly n-doped in the unbiased state. This is rationalized by the natural n-doping of exfoliated transition metal dichalcogenides due to chalcogen atom vacancies as well as extrinsic disorder incurred during (nano)fabrication[49–51]. Despite this, the device performance is not significantly impaired because the effects of unwanted doping are readily mitigated via electrostatic gating. Second, the measured reflectance appears to exceed unity which would imply that there is gain in the structure. In practice, this results from a small normalization error that we attribute to laser power fluctuations and a suboptimal choice of reflection reference in our experiments. Last, we observe intriguing Fabry-Pérot oscillations in the dispersions, which we attribute to self-interference of the guided photons (polaritons) in the heterostructure cavity due to the mismatch between the large non-local mode profile and the finite lateral extent of the cavity (*i.e.*, a finite-size effect). Yet, apart from these deviations, the measured dispersion relations are in excellent agreement with our numerical simulations (Fig. 2c, d).

Next, we extract cross sections from the dispersion relations at $k_x = 0$ μm$^{-1}$ (Fig. 3e). At +25 V, the exciton is suppressed, and the reflectance spectrum contains one near-critically coupled resonance that dips to a minimum of 4% at the operating wavelength $\lambda_0 = 620.5$ nm. Reversing the bias to –25 V neutralizes the monolayer, causing the spectrum to split into two strongly coupled polariton branches and thereby raising the reflectance to 39%. By switching between these two states, our modulator achieves a modulation depth of 9.9 dB ($\Delta R = 35\%$) at the operating wavelength (Fig. 3f) and a change in reflectance up to $\Delta R = 49\%$ at 627.5 nm. To our knowledge, the measured modulation ratio and reflectance contrast are ~5 times greater than previously reported for excitonic metasurfaces at room temperature[30], demonstrating the merits of our hybrid-2D approach.

We additionally assess the modulation frequency response of two more devices in alternating current experiments (Supplementary Fig. S5 and Fig. S3b) and measure a maximum –3 dB bandwidth of $f_{-3dB} = 26$ Hz. While, in principle, the solid-state gating mechanism of excitons allows for very rapid modulation, the low bandwidth observed here can be attributed to the large resistor-capacitor time constant of the electrical circuit owing to the high contact resistance between Au and $WS_2$[52]. We emphasize this is merely a technological rather than a fundamental limitation, as bandwidths exceeding the megahertz regime have already been realized with optimized contact geometries in similar device architectures[29,36].

**Electrically tunable strong coupling**

To gain insight into the voltage-dependence of the fundamental rates governing the metasurface's optical response, we measure the normal-incidence reflectance at gate voltages between ±25 V in intervals of 5 V (Fig. 4a). With increasing voltage, the coupled modes gradually merge into a single resonance, continuously transitioning from strong to weak coupling[26–29,43]. In the following, we quantify the characteristic decay and coupling rates from our measurements by performing a comparative analysis using our RCWA and CMT models.

To accurately model the fabricated device in our RCWA simulations, we narrow down the exact geometry using reflectometry and atomic force microscopy, measure the relative permittivity of CSAR with spectroscopic ellipsometry, and obtain the dielectric functions of the other materials from literature[53–55]. For monolayer $WS_2$, we model the permittivity with a Cauchy-Lorentz oscillator model where we assign an oscillator to each excitonic feature and a constant background term to account for higher-order resonances[4,23,55,56] (see Methods for details). In our analysis, the only remaining fitting parameters are the A-exciton energy and decay rates. We start by fitting the measurements at intrinsic doping (Supplementary Fig. S6a,b) and obtain $E_A$ = 2.00 eV, $\hbar\gamma_{A,r}$ = 2.7 meV and $\hbar\gamma_{A,nr}$ = 26.7 meV, closely matching the values used for our numerical simulations (Fig. 2). In the following, we assume that $\gamma_{A,r}$ is voltage-independent, since it would otherwise over-parametrize the model as $\gamma_{A,r}/\gamma_{A,nr} \to 0$, while noting that this may be a slight oversimplification[4]. Nevertheless, the modulator's voltage-dependent optical response is well-captured by the RCWA model (Fig. 4b), with the simulated reflectance map clearly showing two distinct exciton-polariton branches that gradually merge into a single resonance with increasing $\gamma_{A,nr}$. As such, we continue the fitting procedure to extract $\gamma_{A,nr}$ (and $E_A$) at each voltage (Fig. 4c). We find that, in the strongly n-doped regime (at +25 V), the decay rate is increased fourfold to $\hbar\gamma_{A,nr}$ = 109.4 meV while the resonance energy is redshifted to $E_A$=1.99 eV. We emphasize that the modulator's performance could be improved even further by increasing $\gamma_{A,nr}$ to achieve true critical coupling. The observed redshift is attributed to a doping-induced lowering of the exciton binding energy due to screening of the Coulomb interaction, consistent with earlier work[6]. Regardless, the data is captured exceptionally well by our model, allowing us to directly overlay the fitting results on the simulated reflectance map (Fig. 4b) and confirming the validity of our approach.

Next, we determine the exciton-photon coupling strength (inset of Fig. 4c) by plugging the extracted exciton parameters into our CMT model and fitting it to the experimental data (Supplementary Fig. S6c,d). From this, we estimate a voltage-averaged value of $\hbar g$ = 16.6 ± 1.0 meV and find that the excitonic and photonic modes remain strongly coupled up to a total exciton decay rate of $\gamma_{A,max}$ = 43.3 meV. Notably, the measurement at 0 V is also strongly coupled and appears to be a significant outlier. However, this is attributed to an imperfect measurement sequence resulting in hysteresis due to trapped charges[49]. To gain insight in the gating-induced doping density $n_e(V_G)$ in the $WS_2$ monolayer, we describe our device as a parallel-plate capacitor modified to account for quantum capacitance effects[57,58]. We fit the experimental decay rates by plugging the expression for $n_e$ into the optical rate-equation model proposed in earlier work[30,59]. From the fit, we obtain an estimate of the charge neutrality point, $V_{CNP}$ = −21.3 V, which enables us to calibrate $n_e$ with respect to $V_G$ (Fig. 4c). We estimate carrier concentrations ranging from $n_e \approx -0.4 \cdot 10^{12}$ cm$^{-2}$ at −25 V up to $n_e \approx 7.3 \cdot 10^{12}$ cm$^{-2}$ at +25 V, with the maximum electron density up to which the device remains strongly coupled $n_{e,max} \approx 2.4 \cdot 10^{12}$ cm$^{-2}$ (−5.2 V). Altogether, our results reveal that the metasurface's ability to electrically tune the hybrid-2D system in and out of the strong exciton-photon coupling regime fundamentally stems from doping-induced variations in the nonradiative decay rate of excitons in monolayer $WS_2$.

**Conclusion**

This work introduces a hybrid-2D metasurface platform that combines the merits of a tunable Van der Waals heterostructure cavity with the strong light-matter interaction of a non-local dielectric metasurface to achieve strong and tunable exciton-photon interactions at room temperature, which is leveraged for free-space optical modulation. The platform's simple yet versatile design readily accommodates other materials and is straightforwardly adapted to polarization-independent response and various optical functionalities. More broadly, our platform highlights new opportunities for excitonic 2D metasurfaces for dynamic wavefront manipulation in ultra-compact metadevices.

## Methods

### Sample fabrication

**Lithography of prepatterned substrate.** We prepattern Au contacts on diced 12×12 mm$^2$ Si substrates with 100 nm dry-grown thermal oxide (University Wafer). For this, we spin-coat a 400 nm layer of positive-tone electron-beam (e-beam) resist CSAR AR-P 6200 (AllResist GmbH) using a Süss MicroTec Delta 80 and cure it at 175 °C for 120 seconds. We pattern the contact outlines in the resist using e-beam lithography (Raith Voyager, 50 kV acceleration voltage) with a dose of 152.2 µC/cm², and develop them by soaking for 60 seconds in n-amyl acetate, followed by 7 seconds in o-xylene and 15 seconds in MIBK:IPA 9:1. After development, we rinse the substrates in IPA and blow-dry them with $N_2$ gas. With the mask completed, we deposit a 3 nm adhesion layer of Cr and 100 nm Au using a custom-built thermal evaporator. We perform lift-off by soaking the samples in methoxybenzene (anisole) at 70 °C for 20 minutes. Subsequently, we clean the substrates thoroughly with a base piranha solution (1:1 vol. $NH_3$:$H_2O_2$) at 75 °C for 10 minutes, rinse briefly in IPA, and blow-dry with $N_2$ gas. Finally, we treat the substrates in $O_2$/$N_2$ plasma for 2 minutes at 70 W (Diener Zepto) to complete the prepatterned substrates.

**Heterostructure fabrication.** We obtain hBN and $WS_2$ bulk crystals commercially (HQ Graphene) and manually exfoliate them using Nitto SPV-224 tape. We transfer the exfoliated flakes onto stamps cut from polydimethylsiloxane sheets (Gel-Pak WF-30). Using a confocal WITec α-300 microscope, we identify hBN flakes of suitable thickness with white-light reflectometry and $WS_2$ monolayers with direct photoluminescence imaging using a 405 nm widefield excitation source (Becker & Hickl) and a 550 nm long-pass filter (Thorlabs). After identifying the candidate flakes, we mount a prepatterned substrate in a custom stamping microscope with a heating stage. We align the bottom hBN flake with the back-reflector (gate) and the ground electrode, then bring it into contact. We heat the stack to 60 °C and thermalize it for at least 5 minutes. Subsequently, we peel the stamp slowly from the substrate at 0.1 µm/s to transfer the hBN flake. We thermally anneal the substrate with the hBN in a tube furnace in vacuum (P ≈ $10^{-6}$ – $10^{-7}$ mbar) at 150 °C for 8 hours to promote adhesion and clean the exposed surface. We repeat this process for the $WS_2$ monolayer (annealing at 90 °C for 8 hours) and the top hBN layer (120 °C for 24 hours), completing the heterostructure cavity.

**Lithography of subwavelength grating.** To fabricate the subwavelength gratings, we clean the devices briefly with a 30 second $O_2$ descum (Oxford Instruments Plasma 80). We spin-coat a ~220 nm thick layer of CSAR AR-P 62 e-beam resist and cure it at 150 °C for 120 seconds. Using the e-beam lithography system (Raith Voyager), we write the grating patterns with an electron dose of 130.8 µC/cm². We develop the gratings by soaking the samples for 60 seconds in n-amyl acetate, followed by 7 seconds in o-xylene and 15 seconds in MIBK:IPA 9:1. We rinse the samples in IPA and dry them with $N_2$ gas.

**Electrical connections.** To enable electrical contact and mounting in the optical microscope, we attach one of the samples to a custom printed circuit board (PCB) using double-sided tape. We wire-bond the gate and ground electrodes to the PCB with 25 µm diameter Al wires (West.Bond 7KE), connecting them to two separate pads on the PCB.

### Optical measurements

**Back-focal plane reflection.** We perform in-situ back-focal plane (BFP) reflection measurements with a WITec α300 confocal microscope and an NKT Photonics SuperK laser system as the light source (Supplementary Fig. 2a). We generate linearly polarized monochromatic light (~1 nm bandwidth) with a supercontinuum white-light laser (Extreme EXW-12) and an acousto-optically tunable filter (Select VIS/1X), feeding it into the microscope via a single-mode photonic crystal fiber (Connect FD7-PM). We focus the laser light with a 20x objective (Zeiss EC Epiplan, NA = 0.4) to illuminate the sample which is mounted on the microscope's piezo stage. To switch between TE and TM polarized illumination, we rotate the sample (holder) by a quarter turn. We capture the reflected light with the same objective, pass it through a polarization analyzer and the Bertrand lens, and project the BFP onto a 14-bit CMOS camera (Zeiss Axiocam 705 mono). We sweep the laser across the spectrum in steps of 0.5 nm or 1 nm and record an image of the Fourier plane at each wavelength, adjusting the camera's exposure time to account for spectral variation in laser power. For each measurement, we record two reference measurements: one without a sample mounted to remove dark counts and internal reflections, and another on an exposed gold electrode to normalize the data using the theoretical angle-resolved reflectance of 100 nm gold (simulated with RCWA). For DC gating measurements, we apply a gate bias with a Keithley 2612B SourceMeter, sweeping from 0 to +25 V and then from −5 to −25 V in 5 V steps. We use custom Python scripts to interface with the equipment and automate the experiment.

We extract the $k_x$ and $k_y$ dispersions from the BFP images by determining the BFP center ($k_x = k_y = 0$ μm$^{-1}$) manually and defining the coordinate transformation between image pixels and the lens's NA using the known BFP radius (in pixels). A distinct dust particle visible on all BFP images serves to align the images, ensuring a common center coordinate. We define the $k_x = 0$ and $k_y = 0$ μm$^{-1}$ lines based on the sample's orientation and enhance the signal-to-noise ratio by integrating traces over a small angular region (−3 ≤ θ ≤ 3 mrad, 11 pixels). By repeating this process at each wavelength, we construct line-by-line images of the BFP dispersion to generate the diagrams shown in (Fig. 3c, d). We extract the normal-incidence reflectance spectra (Fig. 3e, 4a) and modulation depth (Fig. 3f) from these diagrams by taking wavelength (energy) traces at $k_x = 0$ μm$^{-1}$.

**Alternating current modulation.** We perform alternating current measurements (Supplementary Fig. 5) using a modified optical microscope setup for BFP measurements (Supplementary Fig. 3b). We feed the tunable laser (NKT) into the microscope (WITec) at another fiber port, adding a lens to achieve collimated wide-field excitation after passing through the 20x objective lens (Zeiss). An Agilent HP 33120A function generator provides the square-wave modulation signal. The modulated beam is collected with the same objective, outcoupled into a fiber and detected with a biased Si photodiode (Thorlabs DET100A2). We connect the photodiode electrically to a source-measure unit (Keithley) and amplify the photogenerated current (responsivity ~ 0.43 A/W) while converting it to voltage (0.5 nA/V) with the Stanford Research Systems (SRS) SR-570 transimpedance amplifier. Finally, we measure the signal using an SR-830 lock-in amplifier (SRS) or a TDS 3032 oscilloscope (Tektronix) synced to the function generator. All measurements are automated using custom Python scripts.

**Spectrally resolved photoluminescence mapping.** We perform photoluminescence (PL) spectroscopy and mapping in the WITec microscope with a 100x objective (Zeiss EC Epiplan-Neofluar, NA = 0.9) in dark-field mode. We use a fiber-coupled 532 nm diode laser to excite a diffraction-limited spot on the sample. The collected PL emission is chromatically dispersed by a WITec UHTS300 SMFC VIS spectrograph (600 lines/mm), and we record the spectra on an Andor Newton EMCCD camera cooled to −60 °C. For spatial mapping, we precisely control the sample position using a piezo-actuated stage and acquire spectra on a rectangular grid with 333 nm spacing.

**Numerical simulations**
**Rigorous coupled-wave analysis.** Numerical simulations are performed using the S$^4$ RCWA package available for Python[60]. We obtain the refractive indices of Au and hBN from literature as tabulated data[53,54], while for CSAR we measure it through spectroscopic ellipsometry (J.A. Woollam VB-400) of the material spin-coated on a Si substrate. For monolayer WS$_2$, we use Cauchy-Lorentz oscillator model to express the permittivity as a set of excitonic oscillators—each characterized by a resonance energy, oscillator strength and total decay rate—plus a constant background term to account for higher-order resonances[4,23,55,56]. We modify the model by expressing the A-exciton's oscillator strength in terms of its radiative decay rate $\gamma_{A,r}$ as $f_A \approx \hbar \gamma_{A,r} \cdot 2\hbar c d^{-1}$, where $\hbar$ is the reduced Planck's constant, $c$ the speed of light, and $d$ = 6.18 Å the thickness of monolayer WS$_2$[61]. To start, the A-exciton energy and linewidth are estimated from preliminary measurements and checked by fitting a spectrally resolved PL map (Supplementary Fig. 2). We narrow down the exact geometry of the fabricated modulator by measuring the thicknesses of the hBN layers using white-light reflectometry in the WITec microscope, and the period, height and width of the CSAR grating using atomic force microscopy (Bruker Dimension FastScan, Fig. 3b). To fit the voltage-dependent reflectance spectra, we initially treat the A-exciton and the geometry as fitting parameters constrained by the experimental uncertainty to account for variations across the sample area. We start by fitting the reflectance at $V_G = -25$V because the clearly distinguishable polariton branches allow for reliable determination of the desired parameters (Supplementary Fig. 6a). In subsequent fits, the geometry is fixed because it does not change with voltage, and $\gamma_{A,r}$ is fixed because it would over-parametrize the model in the exciton-quenched regime.

**Coupled mode theory.** We employ a CMT analysis to evaluate the strong coupling condition and to extract the coupling strength $g$ and polariton dispersions. We define the complex energy of oscillator $j$ as $\tilde{E}_j = E_j - i\hbar\Gamma_j$, with resonance energy $E_j$ and damping rate $\Gamma_j = \gamma/2$, such that the eigenenergies of the coupled excitonic (A) and photonic (GMR) modes are given by:

$$\tilde{E}_\pm = \left(\frac{\tilde{E}_{GMR}+\tilde{E}_A}{2}\right) \pm \sqrt{(\hbar g)^2 - \left(\frac{i(\tilde{E}_{GMR}-\tilde{E}_A)}{2}\right)^2}$$

The Rabi splitting is defined as $\Omega_R = \tilde{E}_+ - \tilde{E}_-$ and the strong coupling criterion as $\Re(\Omega_R) > \hbar\Gamma_{GMR} + \hbar\Gamma_A$. In evaluating this condition, we account for doping-dependent variations in $E_A$ by invoking $E_{GMR} = E_A$ such that the Rabi splitting energy reduces to:

$$\Omega_R = \hbar\sqrt{4g^2 - (\Gamma_{GMR} - \Gamma_A)^2}.$$

To fit the experimental reflectance (Supplementary Fig. 6c, d), we model the non-resonant background reflection using RCWA simulations for TM-polarized illumination (Fig. 2b). We plug the extracted values of $\tilde{E}_A$ and $\tilde{E}_{GMR}$ from our RCWA fits (Supplementary Fig. 6a, b) into the CMT model, leaving only $g$ and two normalization terms $I_A$ and $I_{GMR}$ as free fitting parameters, which we obtain via a least-squares fitting routine. The uncertainty in $g$ is quantified as the mean relative error between data and fit (inset of Fig. 4c). The numerical analyses are performed using custom Python scripts.

**Carrier density calculations.** For calculations of the 2D electron density $n_e$ as function of the applied gate bias, we model our device as a parallel-plate capacitor. We modify the well-known expression for the parallel-plate capacitance, $C_{geo} = \varepsilon_0\varepsilon_{hBN}/d_{hBN}$, to account for the quantum capacitance[57,58], $C_q = e^2 g_s g_v m^*/(\pi\hbar^2)$, such that the overall capacitance per unit area is given as $C = (1/C_{geo} + 1/C_q)^{-1}$. We use a custom Python script to calculate $n_e$ by plugging the measured hBN thickness $d_{hBN}$ = 117 nm and its static permittivity[62] $\varepsilon_{hBN} \approx 3.4$, together with values for the monolayer WS$_2$ bandgap energy[63] $E_{bg} \approx 2.15$ eV, valley degeneracy $g_v$ = 2, spin degeneracy $g_s$ = 2, and effective electron mass[64] $m^* \approx 0.34 m_e$. We obtain an estimate of charge neutrality point by fitting $\gamma_{A,nr}(n_e)$ using an optical rate-equation model[30,59] (Fig. 4c).

## References

1. Geim, A. K. & Grigorieva, I. V. Van der Waals heterostructures. *Nature* **499**, 419–425 (2013).
2. K. S. Novoselov, A. K. Geim, S. V. Morozov, D. Jiang, Y. Zhang, S. V. Dubonos, I. V. G. and A. A. F. Electric Field Effect in Atomically Thin Carbon Films. **306**, 666–669 (2004).
3. Chernikov, A. *et al.* Exciton Binding Energy and Nonhydrogenic Rydberg Series in Monolayer WS$_2$. *Phys. Rev. Lett.* **113**, 076802 (2014).
4. Li, M., Biswas, S., Hail, C. U. & Atwater, H. A. Refractive Index Modulation in Monolayer Molybdenum Diselenide. *Nano Lett.* **21**, 7602–7608 (2021).
5. Yu, Y. *et al.* Giant Gating Tunability of Optical Refractive Index in Transition Metal Dichalcogenide Monolayers. *Nano Lett.* **17**, 3613–3618 (2017).
6. Chernikov, A. *et al.* Electrical Tuning of Exciton Binding Energies in Monolayer WS$_2$. *Phys. Rev. Lett.* **115**, 126802 (2015).
7. Ross, J. S. *et al.* Electrical control of neutral and charged excitons in a monolayer semiconductor. *Nat. Commun.* **4**, 1474 (2013).
8. Zhou, Y. *et al.* Controlling Excitons in an Atomically Thin Membrane with a Mirror. *Phys. Rev. Lett.* **124**, 027401 (2020).
9. Kim, I. *et al.* Nanophotonics for light detection and ranging technology. *Nat. Nanotechnol.* **16**, 508–524 (2021).
10. Miller, D. A. B. Attojoule Optoelectronics for Low-Energy Information Processing and Communications. *J. Light. Technol.* **35**, 346–396 (2017).
11. Zheludev, N. I. & Kivshar, Y. S. From metamaterials to metadevices. *Nat. Mater.* **11**, 917–924 (2012).
12. Kamali, S. M., Arbabi, E., Arbabi, A. & Faraon, A. A review of dielectric optical metasurfaces for wavefront control. *Nanophotonics* **7**, 1041–1068 (2018).
13. Shaltout, A. M., Shalaev, V. M. & Brongersma, M. L. Spatiotemporal light control with active metasurfaces. *Science* **364**, eaat3100 (2019).

14. Overvig, A. & Alù, A. Diffractive Nonlocal Metasurfaces. *Laser Photonics Rev.* **16**, 2100633 (2022).
15. Gu, T., Kim, H. J., Rivero-Baleine, C. & Hu, J. Reconfigurable metasurfaces towards commercial success. *Nat. Photonics* **17**, 48–58 (2023).
16. Kuznetsov, A. I. *et al.* Roadmap for Optical Metasurfaces. *ACS Photonics* **11**, 816–865 (2024).
17. Shirmanesh, G. K., Sokhoyan, R., Wu, P. C. & Atwater, H. A. Electro-optically Tunable Multifunctional Metasurfaces. *ACS Nano* **14**, 6912–6920 (2020).
18. Benea-Chelmus, I. C. *et al.* Gigahertz free-space electro-optic modulators based on Mie resonances. *Nat. Commun.* **13**, 3170 (2022).
19. Sherrott, M. C. *et al.* Experimental Demonstration of >230° Phase Modulation in Gate-Tunable Graphene-Gold Reconfigurable Mid-Infrared Metasurfaces. *Nano Lett.* **17**, 3027–3034 (2017).
20. Holsteen, A. L., Raza, S., Fan, P., Kik, P. G. & Brongersma, M. L. Purcell effect for active tuning of light scattering from semiconductor optical antennas. *Science* **358**, 1407–1410 (2017).
21. Pitanti, A., Da Prato, G., Biasiol, G., Tredicucci, A. & Zanotto, S. Gigahertz Modulation of a Fully Dielectric Nonlocal Metasurface. *Adv. Opt. Mater.* **12**, 2401283 (2024).
22. van de Groep, J. *et al.* Exciton resonance tuning of an atomically thin lens. *Nat. Photonics* **14**, 426–430 (2020).
23. Guarneri, L. *et al.* Temperature-Dependent Excitonic Light Manipulation with Atomically Thin Optical Elements. *Nano Lett.* **24**, 6240–6246 (2024).
24. Epstein, I. *et al.* Near-unity light absorption in a monolayer $WS_2$ van der Waals heterostructure cavity. *Nano Lett.* **20**, 3545–3552 (2020).
25. Li, B. *et al.* Single-Nanoparticle Plasmonic Electro-optic Modulator Based on $MoS_2$ Monolayers. *ACS Nano* **11**, 9720–9727 (2017).
26. Lee, B. *et al.* Electrical Tuning of Exciton–Plasmon Polariton Coupling in Monolayer $MoS_2$ Integrated with Plasmonic Nanoantenna Lattice. *Nano Lett.* **17**, 4541–4547 (2017).
27. Zhang, L., Gogna, R., Burg, W., Tutuc, E. & Deng, H. Photonic-crystal exciton-polaritons in monolayer semiconductors. *Nat. Commun.* **9**, 713 (2018).
28. Fernandez, H. A., Withers, F., Russo, S. & Barnes, W. L. Electrically Tuneable Exciton-Polaritons through Free Electron Doping in Monolayer $WS_2$ Microcavities. *Adv. Opt. Mater.* **7**, 1900484 (2019).
29. Dibos, A. M. *et al.* Electrically Tunable Exciton–Plasmon Coupling in a $WSe_2$ Monolayer Embedded in a Plasmonic Crystal Cavity. *Nano Lett.* **19**, 3543–3547 (2019).
30. Li, Q. *et al.* A Purcell-enabled monolayer semiconductor free-space optical modulator. *Nat. Photonics* **17**, 897–903 (2023).
31. Song, J.-G. *et al.* Effect of $Al_2O_3$ Deposition on Performance of Top-Gated Monolayer $MoS_2$-Based Field Effect Transistor. *ACS Appl. Mater. Interfaces* **8**, 28130–28135 (2016).
32. Moody, G. *et al.* Intrinsic homogeneous linewidth and broadening mechanisms of excitons in monolayer transition metal dichalcogenides. *Nat. Commun.* **6**, 8315 (2015).
33. Cadiz, F. *et al.* Excitonic linewidth approaching the homogeneous limit in $MoS_2$-based van der Waals heterostructures. *Phys. Rev. X* **7**, 021026 (2017).
34. Back, P., Zeytinoglu, S., Ijaz, A., Kroner, M. & Imamoğlu, A. Realization of an Electrically Tunable Narrow-Bandwidth Atomically Thin Mirror Using Monolayer $MoSe_2$. *Phys. Rev. Lett.* **120**, 03740 (2018).
35. Scuri, G. *et al.* Large Excitonic Reflectivity of Monolayer $MoSe_2$ Encapsulated in Hexagonal Boron Nitride. *Phys. Rev. Lett.* **120**, 037402 (2018).
36. Andersen, T. I. *et al.* Beam steering at the nanosecond time scale with an atomically thin reflector. *Nat. Commun.* **13**, 3431 (2022).
37. Li, M., Hail, C. U., Biswas, S. & Atwater, H. A. Excitonic Beam Steering in an Active van der Waals Metasurface. *Nano Lett.* **23**, 2771–2777 (2023).
38. Wang, Z., He, L., Kim, B. & Zhen, B. Control over cavity exciton polaritons in monolayer semiconductors. Preprint at http://arxiv.org/abs/2311.03750 (2023).
39. Sortino, L. *et al.* Van der Waals heterostructure metasurfaces: atomic-layer assembly of ultrathin optical cavities. Preprint at https://doi.org/10.48550/arXiv.2407.16480 (2024).
40. Wang, S. S. & Magnusson, R. Theory and applications of guided-mode resonance filters. *Appl. Opt.* **32**, 2606 (1993).

41. Mak, K. F. *et al.* Tightly bound trions in monolayer MoS$_2$. *Nat. Mater.* **12**, 207–211 (2013).
42. Das, S., Wang, Y., Dai, Y., Li, S. & Sun, Z. Ultrafast transient sub-bandgap absorption of monolayer MoS$_2$. *Light Sci. Appl.* **10**, 27 (2021).
43. Chakraborty, B. *et al.* Control of Strong Light–Matter Interaction in Monolayer WS$_2$ through Electric Field Gating. *Nano Lett.* **18**, 6455–6460 (2018).
44. Liu, X. *et al.* Strong light–matter coupling in two-dimensional atomic crystals. *Nat. Photonics* **9**, 30–34 (2015).
45. Dovzhenko, D. S., Ryabchuk, S. V., Rakovich, Y. P. & Nabiev, I. R. Light-matter interaction in the strong coupling regime: Configurations, conditions, and applications. *Nanoscale* **10**, 3589–3605 (2018).
46. Fan, S., Suh, W. & Joannopoulos, J. D. Temporal coupled-mode theory for the Fano resonance in optical resonators. *J. Opt. Soc. Am. A* **20**, 569 (2003).
47. Yu, Y. & Cao, L. Coupled leaky mode theory for light absorption in 2D, 1D, and 0D semiconductor nanostructures. *Opt. Express* **20**, 13847 (2012).
48. Castellanos-Gomez, A. *et al.* Deterministic transfer of two-dimensional materials by all-dry viscoelastic stamping. *2D Mater.* **1**, 011002 (2014).
49. Mitta, S. B. *et al.* Electrical characterization of 2D materials-based field-effect transistors. *2D Mater.* **8**, 012002 (2021).
50. Rhodes, D., Chae, S. H., Ribeiro-Palau, R. & Hone, J. Disorder in van der Waals heterostructures of 2D materials. *Nat. Mater.* **18**, 541–549 (2019).
51. Daus, A. *et al.* High-performance flexible nanoscale transistors based on transition metal dichalcogenides. *Nat. Electron.* **4**, 495–501 (2021).
52. Allain, A., Kang, J., Banerjee, K. & Kis, A. Electrical contacts to two-dimensional semiconductors. *Nat. Mater.* **14**, 1195–1205 (2015).
53. Yakubovsky, D. I., Arsenin, A. V., Stebunov, Y. V., Fedyanin, D. Yu. & Volkov, V. S. Optical constants and structural properties of thin gold films. *Opt. Express* **25**, 25574 (2017).
54. Lee, S. Y., Jeong, T. Y., Jung, S. & Yee, K. J. Refractive Index Dispersion of Hexagonal Boron Nitride in the Visible and Near-Infrared. *Phys. Status Solidi B* **256**, 1800417 (2019).
55. Li, Y. *et al.* Measurement of the optical dielectric function of monolayer transition-metal dichalcogenides: MoS$_2$, MoSe$_2$, WS$_2$, and WSe$_2$. *Phys. Rev. B* **90**, 205422 (2014).
56. Hsu, C. *et al.* Thickness-Dependent Refractive Index of 1L, 2L, and 3L MoS$_2$, MoSe$_2$, WS$_2$, and WSe$_2$. *Adv. Opt. Mater.* **7**, 1900239 (2019).
57. Brumme, T., Calandra, M. & Mauri, F. First-principles theory of field-effect doping in transition-metal dichalcogenides: Structural properties, electronic structure, Hall coefficient, and electrical conductivity. *Phys. Rev. B* **91**, 155436 (2015).
58. Ma, N. & Jena, D. Carrier statistics and quantum capacitance effects on mobility extraction in two-dimensional crystal semiconductor field-effect transistors. *2D Mater.* **2**, 015003 (2015).
59. Lien, D. H. *et al.* Electrical suppression of all nonradiative recombination pathways in monolayer semiconductors. *Science* **364**, 468–471 (2019).
60. Liu, V. & Fan, S. S$^4$: A free electromagnetic solver for layered periodic structures. *Comput. Phys. Commun.* **183**, 2233–2244 (2012).
61. Wilson, J. A. & Yoffe, A. D. The transition metal dichalcogenides discussion and interpretation of the observed optical, electrical and structural properties. *Adv. Phys.* **18**, 193–335 (1969).
62. Pierret, A. *et al.* Dielectric permittivity, conductivity and breakdown field of hexagonal boron nitride. *Mater. Res. Express* **9**, 065901 (2022).
63. Roy, S. & Bermel, P. Electronic and optical properties of ultra-thin 2D tungsten disulfide for photovoltaic applications. *Sol. Energy Mater. Sol. Cells* **174**, 370–379 (2018).
64. Liu, L., Kumar, S. B., Ouyang, Y. & Guo, J. Performance Limits of Monolayer Transition Metal Dichalcogenide Transistors. *IEEE Trans. Electron Devices* **58**, 3042–3047 (2011).

**Acknowledgements**
This work was funded by a Vidi grant (VI.Vidi.203.027) from the Dutch National Science Foundation (NWO). We express our gratitude to Thomas Bauer and Sander A. Mann for the assistance and insightful discussions.

**Author contributions**
T.H. and J.v.d.G. conceived the concept behind this research. T.H. fabricated the samples, performed the simulations and measurements. T.H. and J.v.d.G. performed the data analysis and wrote the manuscript.

**Competing interests**
The authors declare no competing interests.

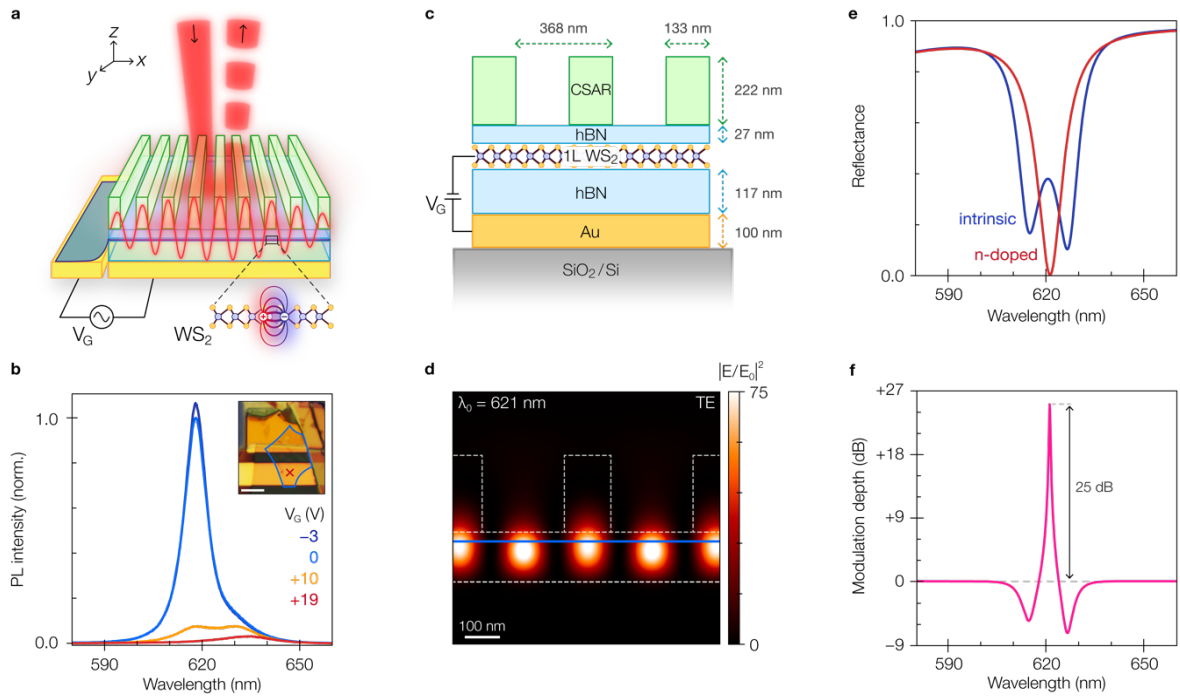

**Figure 1: Hybrid-2D optical modulator. (a)** Illustration of the concept: the electrical tunability of a monolayer 2D semiconductor in a heterostructure cavity is combined with the strong light-matter interaction of a dielectric non-local metasurface to achieve strong intensity modulation of the reflected beam. **(b)** Experimentally measured photoluminescence (PL) intensity for a bare heterostructure cavity as a function of gate voltage (color). The inset shows a micrograph of the heterostructure, with the monolayer outlined in blue and the measurement position indicated by the red cross. Scalebar: 5 μm. **(c)** Schematic of the detailed geometry of the modulator design (not to scale). **(d)** Electric field intensity enhancement for a transverse electric (TE) polarized normal-incident wave at free-space wavelength $\lambda_0 = 621$ nm. Blue line represents the monolayer. **(e)** Calculated reflectance spectra for the device with an intrinsic ($R_i$, blue) and n-doped ($R_n$, red) monolayer. **(f)** The corresponding modulation depth ($10 \cdot \log_{10}(R_i/R_n)$, magenta) extracted from (e).

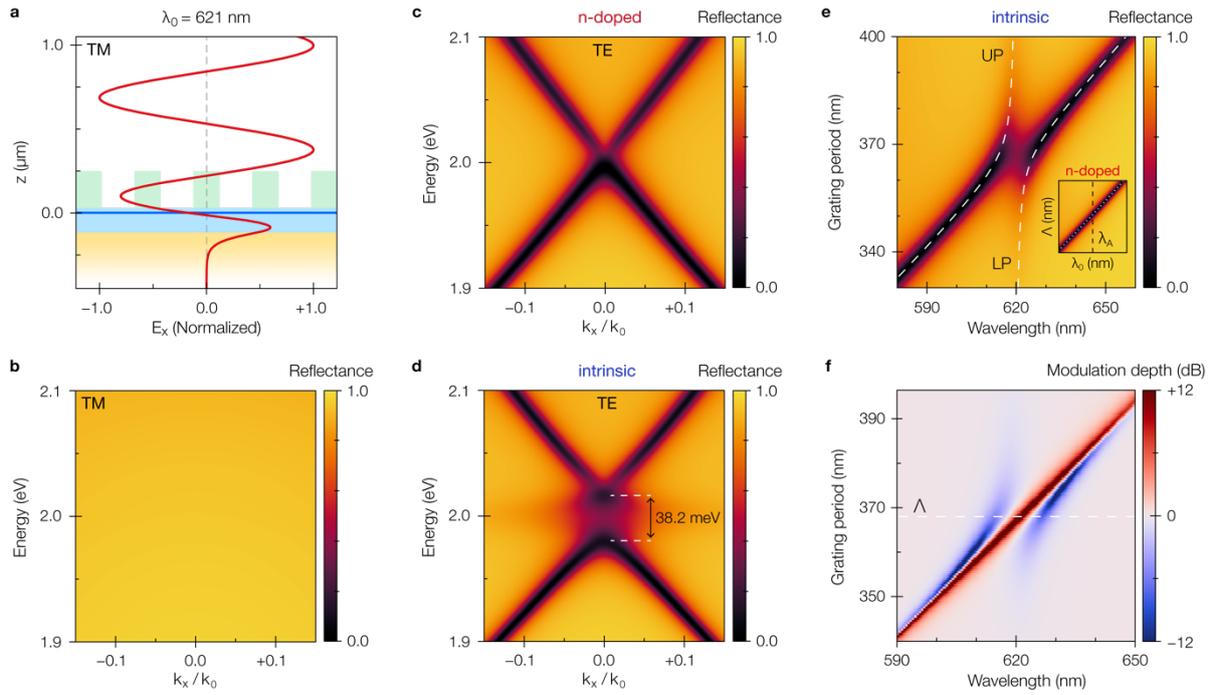

**Figure 2: Nanophotonic design of the hybrid-2D metasurface. (a)** Standing wave profile of the electric field under normal-incidence transverse magnetic (TM)-polarized illumination showing minimal field overlap with the monolayer (blue line at $z$ = 0 μm). **(b-d)** Numerically simulated angular dispersion of the designed modulator under TM (b) and TE-polarized illumination for monolayer $WS_2$ with strong n-type doping (c) and intrinsic doping (d). In the latter case, the A-exciton is strongly coupled to the GMR, resulting in the formation of a 38.2 meV gap (indicated) at the exciton energy. **(e)** Grating period dependence of the reflectance at intrinsic doping and strong n-type doping (inset). The calculated upper (UP) and lower polariton (LP) branches of the coupled mode analysis are represented by the white dashed lines. In the inset, the A-exciton wavelength $\lambda_A$ is represented by the dashed line (black) and the GMR dispersion by the dotted line (white). **(f)** Modulation depth as function of grating period, with the designed period $\Lambda$ indicated (dashed line). The color scale is capped at ±12 dB for visibility.

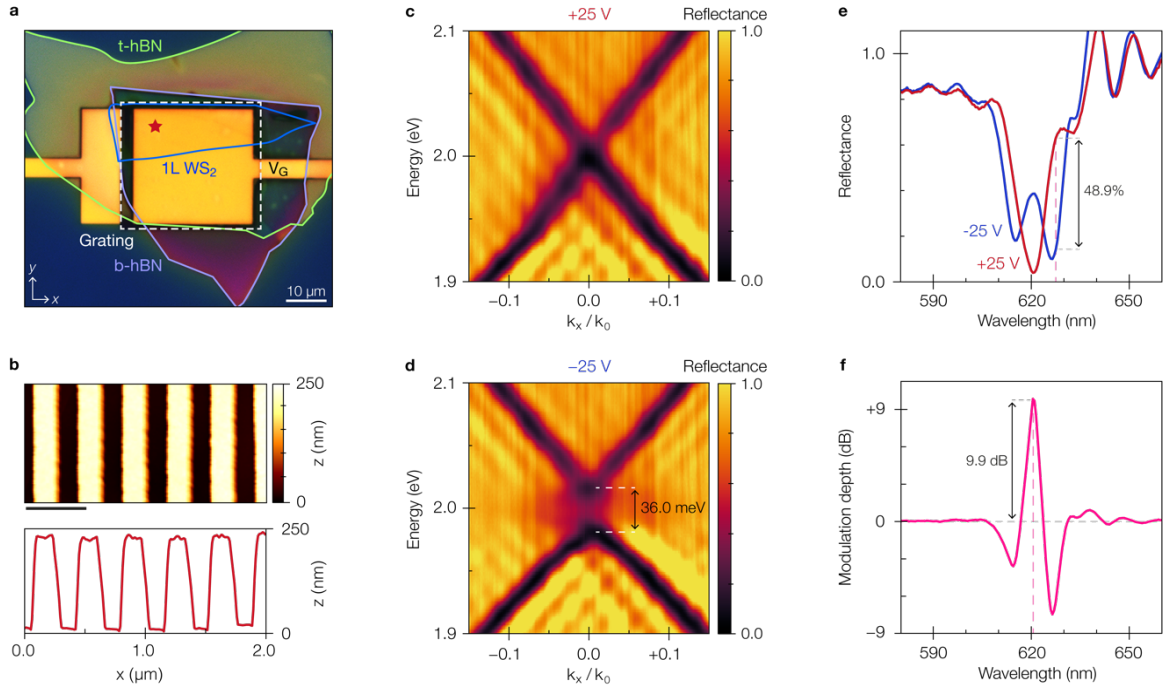

**Figure 3: Free-space optical modulation via electrostatic gating of a hybrid-2D metasurface. (a)** Optical microscope image of the hybrid-2D metasurface: a heterostructure comprised of monolayer WS$_2$ (blue) encapsulated by bottom (purple) and top (green) hBN layers is placed on a gold back-reflector, which doubles as the gate electrode ($V_G$), and integrated with a dielectric grating (white, dashed). The red star indicates the position of the **(b)** atomic force micrograph (top) and horizontal cross-section through the center (bottom). Scalebar: 500 nm. **(c)** Experimental back-focal plane imaging spectroscopy of the angular dispersion at strong n-type doping (+25 V, red) and **(d)** intrinsic doping (−25 V, blue), with the latter exhibiting a Rabi splitting energy of $\Omega_R = 36.0$ meV. **(e)** Normal-incidence TE-polarized reflectance spectra extracted from (c) and (d) at $k_x/k_0 = 0$, with a maximum reflectance contrast of 48.9% at $\lambda_0 = 627.5$ nm indicated. **(f)** Modulation depth obtained from the spectra in (e), revealing 9.9 dB peak signal modulation at $\lambda_0 = 620.5$ nm.

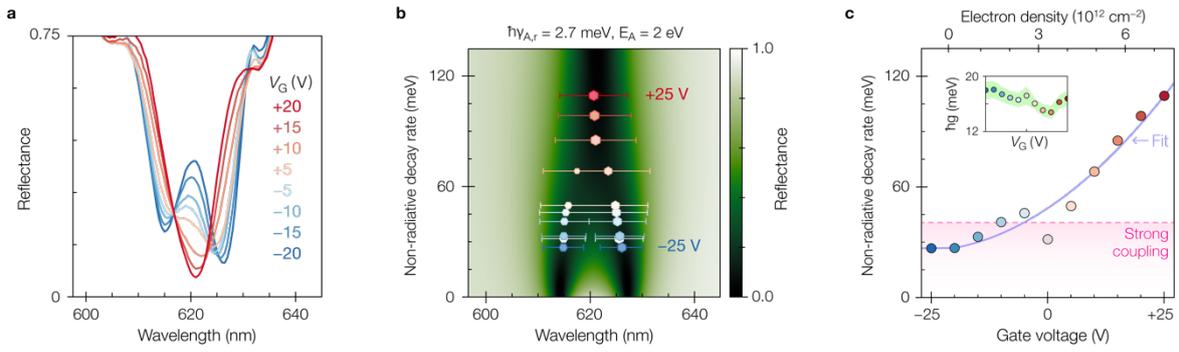

**Figure 4: Tunable strong coupling via electrostatic gating. (a)** Normal-incidence TE-polarized reflectance as function of gate voltage (color). The spectra measured at $V_G = 0$ V, $\pm 25$ V are omitted for clarity. **(b)** Spectral reflectance versus nonradiative decay rate of excitons in monolayer $WS_2$, simulated with the RCWA model for $\hbar\gamma_{A,r} = 2.7$ meV and $E_A = 2$ eV. The colored datapoints represent the fitted energies, relative amplitudes (position and size of hexagons, respectively) and linewidths (width of bars) of the coupled modes. **(c)** Exciton nonradiative decay rate as function of gate voltage (electron density) obtained from the RCWA fits (colored datapoints). Purple curve indicates the fitted optical rate equation, pink-shaded region represents the decay rates at which the system is strongly coupled. The coupling rates $\hbar g$ obtained from the CMT analysis are shown in the inset (fitting uncertainty in green).

# Supporting Information

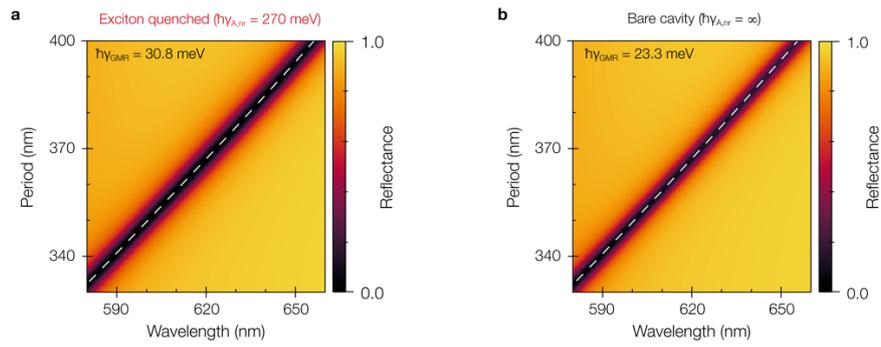

**Supplementary Figure S1: Comparison of the modulator in the exciton-quenched state and the bare cavity. (a)** Numerically simulated period dependence of the normal-incidence TE-polarized reflectance of the designed modulator in the exciton-quenched state, and **(b)** of the corresponding bare cavity. The fitted dispersion is overlaid on the colormaps (dashed) and the extracted damping rate of the guided-mode resonance (GMR) is indicated.

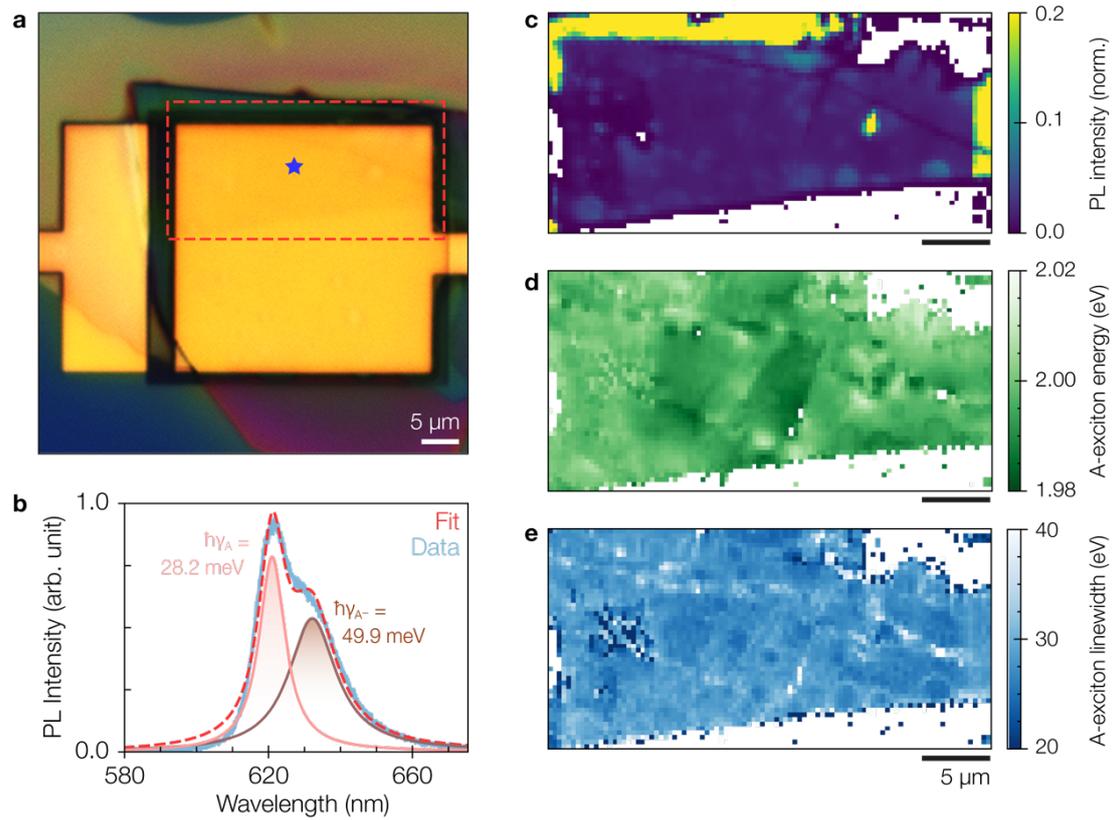

**Supplementary Figure S2: Photoluminescence characterization.** (a) Brightfield microscope image the fabricated modulator. (b) Characteristic PL spectrum (light blue) obtained by averaging a 1 μm² spot indicated by the blue star in (a). The data is fitted with a double Lorentzian function (red, dashed) and separated into A-exciton (pink) and negatively charged trion (A⁻, brown) contributions. (c) Spatially-resolved photoluminescence (PL) maps of the region (red, dashed) outlined in (a), showing the integrated intensity, (d) the fitted A-exciton energy, and (e) the fitted linewidth (total decay rate). Scalebars: 5 μm.

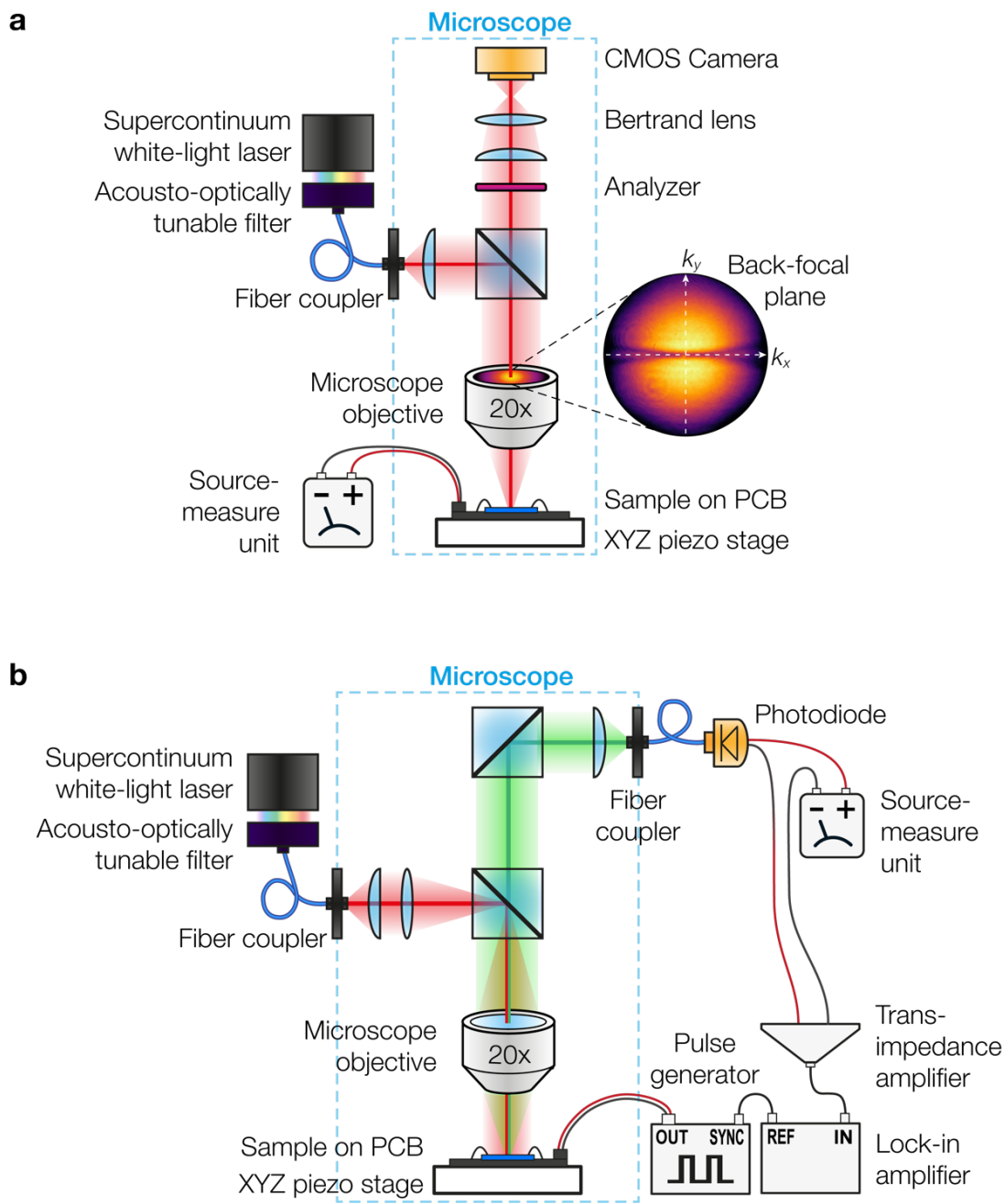

**Supplementary Figure S3: Measurement setups for DC and AC modulation experiments. (a)** Back-focal plane imaging setup used for angle-resolved reflectance spectroscopy and DC modulation experiments. **(b)** Normal-incidence, spectrally resolved AC reflectance modulation setup.

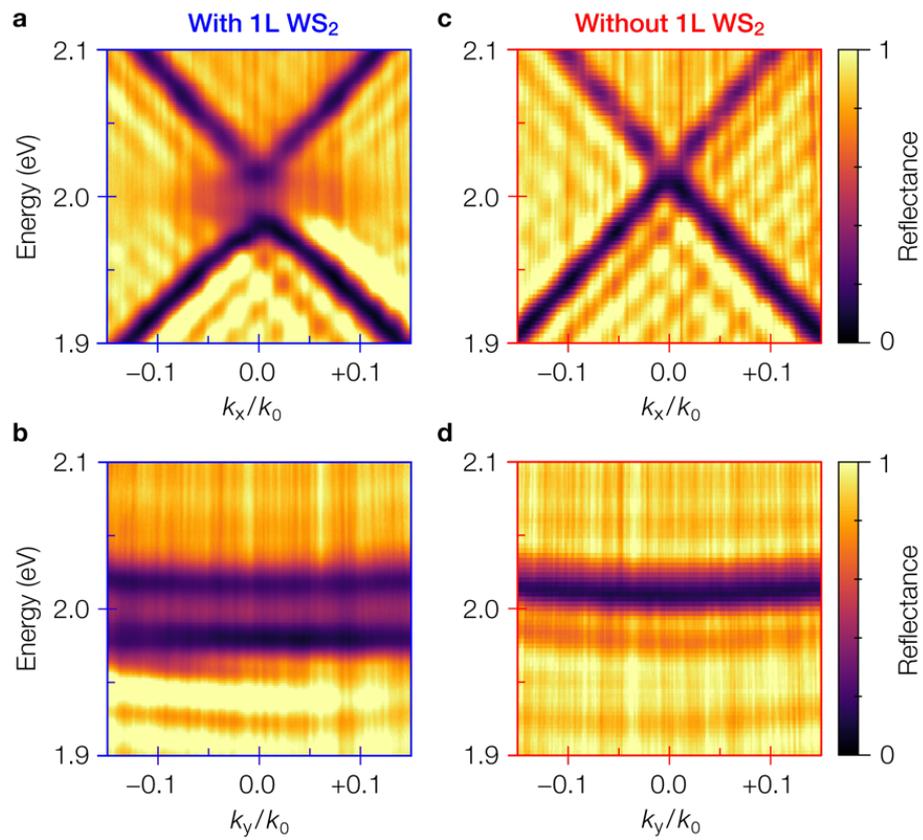

**Supplementary Figure S4: Experimental dispersion of the cavity with and without monolayer WS$_2$.** (a) $k_x$ and (b) $k_y$ dispersion of the fabricated modulator under TE-polarized illumination. (c) $k_x$ and (d) $k_y$ dispersion of the empty cavity, measured on a patch of the device that lacks the WS$_2$ monolayer.

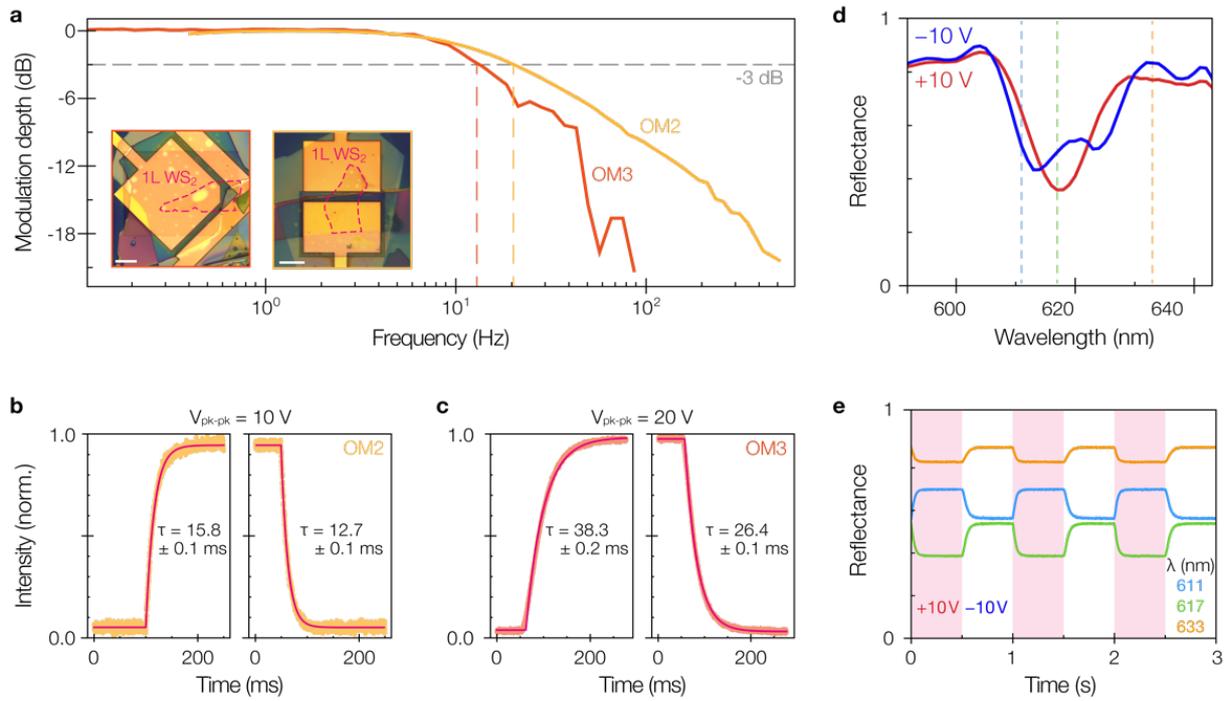

**Supplementary Figure S5: AC modulation characteristics of two hybrid-2D modulators. (a)** Modulation bandwidth of devices OM2 (yellow) at $\lambda_0 = 631$ nm with 3 dB cut-off at $f_{-3dB} = 12.8$ Hz and OM3 (orange) at $\lambda_0 = 633$ nm with $f_{-3dB} = 20.2$ Hz. Inset shows optical micrographs of OM2 and OM3 with monolayer (1L) WS$_2$ outlined in magenta. Scalebars: 10 μm. **(b, c)** Corresponding time traces of the rise (left) and fall (right) of the reflected intensity in response to a 1 Hz square-wave modulation signal for devices OM2 (b) and OM3 (c). The rise and fall times obtained by fitting (magenta) and the peak-to-peak driving voltage $V_{pk-pk}$ are also shown. **(e)** DC reflectance spectra of device OM3 at +10 V (red) and −10 V (blue) and **(e)** reflectance as function of time in response to a 1 Hz pulse at $\lambda_0 = 611$ nm (light blue), 617 nm (green) and 633 nm (orange). Colors corresponding to dotted lines in (d).

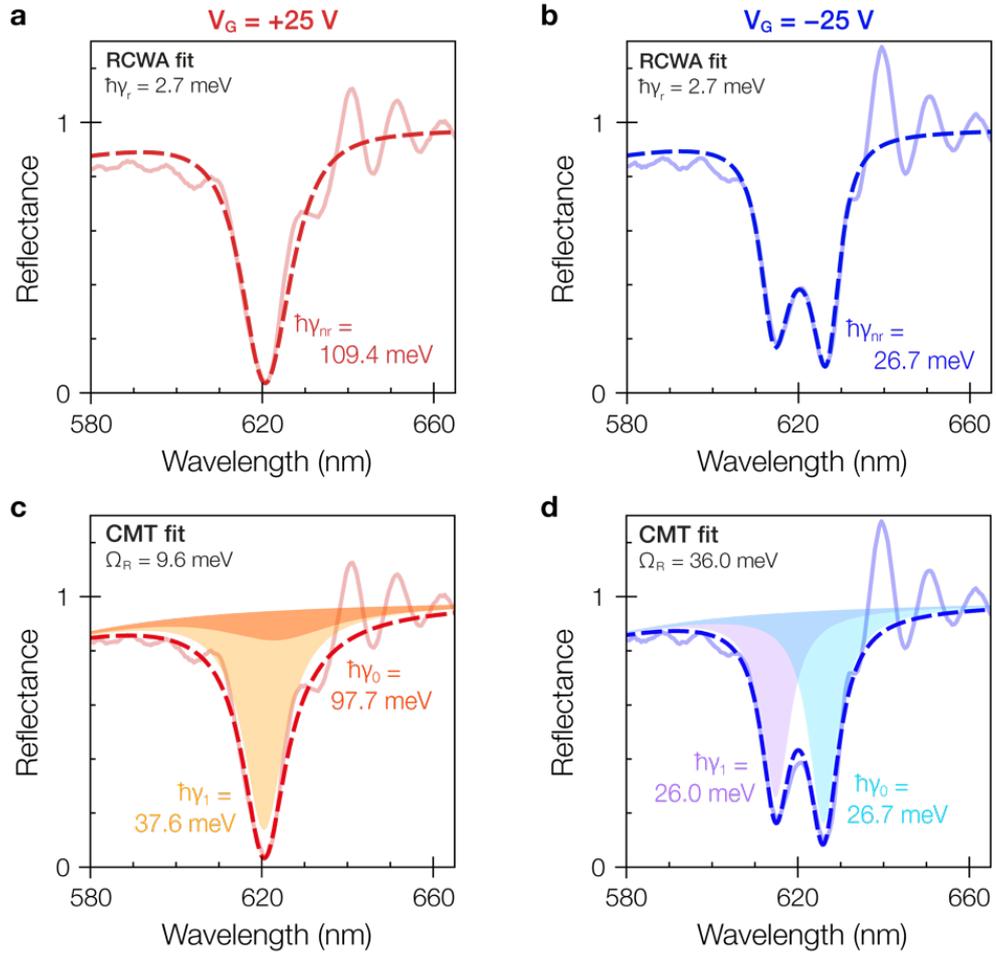

**Supplementary Figure S6: Parameter extraction via reflectance fitting.** (a) Fit (dashed) of the normal-incidence reflectance performed with rigorous coupled-wave analysis (RCWA) for an n-doped (+25 V), and (b) a neutral (−25V) monolayer, used to extract the A-exciton energy plus the radiative and nonradiative decay rates, $\gamma_r$ and $\gamma_{nr}$, respectively. (c) Fit (dashed) performed with coupled-mode theory (CMT) for an n-doped, and (d) intrinsic monolayer, to extract the exciton-photon coupling strength as well as the energies and linewidths ($\gamma_0$ and $\gamma_1$) of the coupled modes. The CMT fits are offset by a non-resonant background reflectance obtained through RCWA simulations for TM-polarized illumination.